# SELF-AFFINITY IN FINANCIAL ASSET RETURNS

**John Goddard**+

*Bangor University*

**Enrico Onali**

*Bangor University*

## Abstract

We test for departures from normal and independent and identically distributed (NIID) returns, when returns under the alternative hypothesis are self-affine. Self-affine returns are either fractionally integrated and long-range dependent, or drawn randomly from an L-stable distribution with infinite higher-order moments. The finite sample performance of estimators of the two forms of self-affinity is explored in a simulation study which demonstrates that, unlike rescaled range analysis and other conventional estimation methods, the variant of fluctuation analysis that considers finite sample moments only is able to identify either form of self-affinity. However, when returns are self-affine and long-range dependent under the alternative hypothesis, rescaled range analysis has greater power than fluctuation analysis. The finite-sample properties of the estimators when returns exhibit either form of self-affinity can be exploited to determine the source of self-affinity in empirical returns data. The techniques are illustrated by means of an analysis of the fractal properties of the daily logarithmic returns for the indices of 11 stock markets.

+ j.goddard@bangor.ac.uk

# SELF-AFFINITY IN FINANCIAL ASSET RETURNS

## 1. Introduction

Long-range dependence and stable laws in returns have been investigated in the econometrics literature for several decades. Long-range dependence implies a hyperbolic decay of the autocorrelation function in the time domain (Banerjee and Urga, 2005). Stable laws accommodate departures from normality and the related central-limit theorem for independent and identically distributed variables (Levy, 1925). Following the pioneering work of Mandelbrot (1963, 1967, 1971), models that accommodate long-range dependence and stable laws have been employed to describe stock market behaviour. These models represent an application of fractal mathematics to financial economics, a topic that has attracted widespread interest in recent years.[1]

A fractal exhibits the properties of self-similarity or scale invariance. It is suggested by Mandelbrot (1977) that stock returns may exhibit the weaker property of self-affinity. After the application of a suitable rescaling transformation, which takes the form of a single non-random contraction dependent upon the time scale only, a self-affine returns series exhibits the property of self-similarity. A self-affine returns series has the same distributional properties (after rescaling) when returns are measured at any frequency, and is said to be unifractal or monofractal.

Conventional finance literature assumes that logarithmic returns are normal (Gaussian), independent and identically distributed (NIID), and log prices follow random walks (Fama, 1970). Departures from the NIID assumption invalidate several asset pricing models and statistical tools commonly employed in finance, such as the Capital Asset Pricing Model (Sharpe, 1964; Lintner, 1965), and Black-Scholes' (1972, 1973) model of option pricing.

Two classes of process, in which returns are either non-independent or non-Gaussian or both, embody the properties of self-affinity and unifractality (Mandelbrot, Fisher and Calvet, 1997; Cont and Tankov, 2004). First, if returns are fractionally integrated, the autocorrelation function measured over any time scale exhibits the property of long-range dependence, and the log price series is characterized as Fractional Brownian Motion (FBM).[2] The autoregressive fractionally integrated moving average (ARFIMA) model is the best-known member of the class of fractionally integrated processes. Let $p_t$ denote log price at time t, and let $\Delta^{(n)}p_t = p_t - p_{t-n}$ denote the returns measured over the time scale n. The scaling behaviour of $p_t$ is described by the Hurst exponent (Hurst, 1951), denoted H. For a fractionally integrated process, the Hurst exponent is a simple function of the order of fractional integration. For $0.5<H<1$, the local growth rate of $p_t$ is of order $(\Delta t)^H > (\Delta t)^{0.5}$, because the positive autocorrelation in $\Delta^{(1)}p_t$ creates a tendency for $p_t$ to move further in each period than it does in the case H=0.5, where $\Delta^{(1)}p_t$ is NIID.

Second, the class of probability distributions known as Levy-stable, Pareto-Levy stable or L-stable (Levy, 1925; Mandelbrot, 1963, 1967) includes several heavy-tailed distributions with infinite variance and higher-order moments. For an L-stable process, an incidence of large positive or negative returns measured at the highest frequency creates a tendency for $p_t$ to move further in each period than it does in the NIID case. As before, the local growth rate of $p_t$ is of order $(\Delta t)^H > (\Delta t)^{0.5}$, where the Hurst exponent H is a function of the parameterization of the L-stable distribution. $\Delta^{(n)}p_t$ for n>1 has the same (non-Gaussian) distribution as $\Delta^{(1)}p_t$, and is self-affine and unifractal. The infinite variance property renders the central-limit theorem inapplicable, and there is no convergence towards the Gaussian distribution as $n\rightarrow\infty$.

This paper contributes to two strands of literature, on long-range dependence or fractional integration, and on L-stable distributions. We examine the performance of estimators of the Hurst exponent, in the case where there is long-range dependence (and the distribution of returns is Gaussian), and in the case where the distribution of returns is L-stable (and there is no long-range dependence). Hypothesis tests for departures from the NIID case are developed, based on the application of two widely-used methods for estimating the Hurst exponent, to simulated NIID returns data: rescaled range analysis (RRA),[3] and fluctuation analysis (FA).[4] Both methods are based on an examination of the scaling behaviour of selected sample moments, as the time scale over which returns are measured varies.

The performance of the tests under the alternative hypothesis is examined by evaluating power functions, using simulated self-affine series characterized as either long-range dependent, or L-stable with infinite higher-order moments. Monte Carlo simulations are employed, because the asymptotic properties of the RRA and FA estimators are indeterminate (Fisher, Calvet, and Mandelbrot, 1997; Urga and Banerjee, 2005). In addition, we draw comparisons with the performance of other tests widely employed to estimate long-range dependence (Geweke and Porter-Hudak, 1983; Robinson, 1995), and the characteristic exponent of an L-stable distribution (Pickands, 1975; Hill, 1975; de Haan and Resnick, 1980).

In much of the previous literature, researchers have reported evidence concerning the fractal properties of financial returns series in the form of point estimates of the Hurst exponent, or graphical analysis of scaling behaviour.[5] In the absence of any basis for assessing the statistical significance of possible departures from the NIID case, however, much of this evidence is at best suggestive of the possibility that models based on fractal mathematics might provide a more satisfactory representation of the behaviour of returns than models embodying the NIID assumption. This paper relocates several established but informal procedures within a conventional and formal hypothesis testing framework, enabling conclusions to be drawn based on the standard criteria of statistical inference.

The principal findings are as follows. Tests for departure from the NIID case based on RRA and FA perform well when returns are self-affine and long-range dependent under the alternative hypothesis. In this case, the test based on RRA has greater power than the tests based on the three variants of FA that are considered. However, the test based on RRA performs poorly when returns are self-affine and L-stable with infinite higher-order moments under the alternative hypothesis. In this case, the choice of sample moments over which the FA is computed is crucial: the FA should not consider sample moments whose true values are infinite. As an estimator of the Hurst exponent, the variant of the FA that considers finite sample moments only is unique (among the estimators considered in this paper) in terms of its reliability under both of the long-range dependent and L-stable alternatives to the null hypothesis of NIID returns.

The remainder of the paper is structured as follows. Section 2 examines several aspects of the technical background: the property of self-affinity; Monte Carlo simulation of self-affine series; and estimation methods for the Hurst exponent. Section 3 presents critical values for the statistical tests for departure from the NIID case. Section 4 illustrates the techniques described in the previous sections, using data for 11 stock market indices for the period 1987-2011. Finally, Section 5 summarizes and concludes.

## 2. Technical background

Section 2 describes the technical background. Section 2.1 describes the property of self-affinity when returns are fractionally integrated, and therefore long-range dependent. Section 2.2 describes the property of self-affinity when returns are L-stable with infinite higher-order moments, and independent. Section 2.3 describes the methods used in this paper for Monte Carlo simulation of self-affine returns series. Finally, Section 2.4 describes two methods for estimating the Hurst exponent: rescaled range analysis (RRA), and fluctuation analysis (FA).

### 2.1 The self-affinity property: Fractional Gaussian Noise and ARFIMA

Let $\gamma_0^{(n)} = \mathrm{var}(\Delta p_t^{(n)})$ and $\gamma_k^{(n)} = \mathrm{cov}(\Delta p_t^{(n)}, \Delta p_{t-nk}^{(n)})$ denote the autocovariance function for returns measured over time scale n. A returns series is described as Fractional Gaussian Noise (FGN) if $\gamma_0^{(n)} = n^{2H}\gamma_0^{(1)}$ for any n>1, where H is the Hurst exponent. For FGN, the autocovariance function $\gamma_k^{(1)} = (1/2)[(k+1)^{2H} - 2k^{2H} + (k-1)^{2H}]$ is characterized by a single parameter, H. The decay of the autocovariance function as $k\to\infty$ follows a power law, such that $\gamma_k^{(1)} \to k^{-\beta}L(k)$ for $0<\beta<1$, and L(k) satisfies $L(xk)/L(k) \to 1$ as $k\to\infty$ , for any x>0. FGN exhibits the property of self-affinity.

FGN is one member of a family of fractionally integrated processes (Granger and Joyeux, 1980; Hosking, 1981; Geweke and Porter-Hudak, 1983), which includes ARFIMA(p,d,q)

$$(1 - L)^d \Delta^{(1)} p_t = u_t \qquad [1]$$

In [1], L denotes the lag operator, and $u_t$ is NIID. $\Delta^{(1)} p_t$ may incorporate short-range dependence described by p'th order autoregressive (AR) and q'th order moving average (MA) components, as well as long-range dependence. The parameter d is the order of fractional integration. Asymptotically as $k \to \infty$, the autocovariance function for ARFIMA(0,d,0) satisfies the conditions for self-affinity described above, with H=d+0.5. Accordingly, an ARFIMA(0,d,0) returns series is said to be asymptotically self-affine.

## 2.2 The self-affinity property: L-stable processes

The L-stable class of probability distributions is described by the characteristic function $\phi(t)$, defined as follows:

$$\ln[\phi(t)] = -\sigma |t|^{\alpha} \{1 - i\beta \operatorname{sgn}(t) \tan(\pi\alpha/2)\} + i\mu t \qquad \text{for } \alpha \neq 1$$

$$= -\sigma |t| \{1 + i\beta \operatorname{sgn}(t)(2/\pi) \ln(|t|)\} + i\mu t \qquad \text{for } \alpha = 1 \qquad [2]$$

In [2], $\alpha$ is the characteristic exponent, $\beta$ is the skewness parameter, $\mu$ is the location parameter, $\sigma$ is the scale parameter, and sgn(t) = –1 if t<0, sgn(t)=0 if t=0, sgn(t)=1 if t>0. Gaussian returns are represented by ($\alpha$=2, $\beta$=0); and several fat-tailed distributions with infinite variance and higher-order moments are represented by $\alpha<2$.[6] For $\alpha<2$, the local growth rate of $p_t$ is of order $(\Delta t)^H > (\Delta t)^{0.5}$, where the Hurst exponent is $H = 1/\alpha$.

If $\Delta p_{t-s}^{(1)}$ for s=0,...,n–1 are independent drawings from an L-stable distribution defined in accordance with [2], the log characteristic function of $\Delta p_t^{(n)} = \sum_{s=0}^{n-1} \Delta p_{t-s}^{(1)}$ is n *times* the log characteristic function of $\Delta p_{t-s}^{(1)}$, and can be written $-n\sigma |t|^{\alpha} \{1 - i\beta \operatorname{sgn}(t) \tan(\pi\alpha/2)\} + in\mu t$. The log characteristic function of $\Delta p_t^{(n)}$ therefore has the same characteristic exponent $\alpha$ and the same skewness parameter $\beta$ as the log characteristic function of $\Delta p_{t-s}^{(1)}$ (and scale and location parameters, nσ and nμ respectively, that are larger in absolute value). The correspondence of α and β between the log characteristic functions of $\Delta p_t^{(1)}$ and $\Delta p_t^{(n)}$ implies the returns series is self-affine.

## 2.3 Monte Carlo simulation of self-affine returns series

Section 2.3 describes the methods that are used to simulate self-affine returns series, for the two cases where returns are ARFIMA(0,d,0), and returns are L-stable with infinite higher-order moments. Monte Carlo techniques have been widely employed to construct tests for statistics whose finite-sample properties are difficult to determine analytically (Dwass, 1957; Barnard, 1963; Hope, 1968; Birnbaum, 1974; Dufour, 2006).[7]

Using the Wold decomposition, the moving average representation of the ARFIMA(0,d,0) model [1] is

$$\Delta^{(1)}p_t = (1 - L)^{-d}u_t = \sum_{j=0}^{\infty} \gamma_j L^j u_{t-j} \qquad [3]$$

where $\gamma_0=1$ and $\gamma_j = \prod_{k=1}^{j} (d + k - 1)/j!$ for $j \geq 1$. To simulate an ARFIMA(0,d,0) series $Y_t$ for a sample size of T, let $u_t \sim N(0,1)$ for $t=-4999, ..., T$; and let $Y_t = \sum_{s=0}^{4999} \gamma_s L^s u_t$ for $t=1, ... ,T$.

Chambers et al. (1976) propose a method for generating a simulated series drawn from an L-stable distribution with characteristic function [2]. The following description is based on Weron (1996). Generate two independent random variables $V \sim U(-\pi/2, \pi/2)$, and $W \sim \exp(1)$. Compute

$$X_t = S_{\alpha,\beta} \frac{\sin[\alpha(V + B_{\alpha,\beta})]}{[\cos(V)]^{1/\alpha}} \left\{ \frac{\cos[V - \alpha(V + B_{\alpha,\beta})]}{W} \right\}^{(1-\alpha)/\alpha} ; \quad Y_t = \sigma X_t + \mu \quad \text{if } \alpha \neq 1$$

$$X_t = \frac{2}{\pi}\left[\left(\frac{\pi}{2} + \beta V\right)\tan V - \beta \ln\left(\frac{W \cos V}{(\pi/2) + \beta V}\right)\right]; \quad Y_t = \sigma X_t + (2/\pi)\beta\sigma \ln(\sigma) \quad \text{if } \alpha=1$$

where $B_{\alpha,\beta} = \alpha^{-1}\arctan[\beta \tan(\pi\alpha/2)]$ ; $S_{\alpha,\beta} = [1+\beta^2 \tan^2(\pi\alpha/2)]^{1/(2\alpha)}$ [4]

The simulated series $Y_t$ has the characteristic function [2], with parameters $\alpha$, $\beta$, $\mu$ and $\sigma$.

## 2.4 Estimation of the Hurst exponent

Section 2.4 describes two estimation methods for the Hurst exponent: rescaled range analysis (RRA), and fluctuation analysis (FA); and cites some alternative estimation methods for the order of fractional integration of a fractionally integrated series, and for the characteristic exponent (or tail parameter) of an L-stable process.

***Rescaled range analysis***

Estimation of the Hurst exponent for a returns series denoted $\{z_t\}$ using RRA proceeds as follows. Starting from the first observation, subdivide the sample period T into M contiguous subperiods labelled m=1,...,M, each containing n observations, and compute the following:

$$\mu_m = n^{-1} \sum_{t=(m-1)n+1}^{mn} z_t \; ; \qquad S_m = \sqrt{n^{-1} \sum_{t=(m-1)n+1}^{mn} (z_t - \mu_m)^2} \qquad \text{for } m = 1, \dots, M$$

$$x_t = \sum_{s=(m-1)n+1}^{t} (z_s - \mu_m) \qquad \text{for } t=(m-1)n+1, \dots, mn-1; \qquad x_{mn} = 0$$

$$R_m = \max_{t\in m}(x_t) - \min_{t\in m}(x_t) \qquad [5]$$

If Mn<T, the expressions in [5] are also calculated with the subdivision starting from the L+1th observation, where L = T–nM. A second set of M calculated values of $R_m$ and $S_m$ is obtained, indexed m=M+1,...,2M.[8] The R/S statistic for time scale n is

$$(R/S)_n = (2M)^{-1} \sum_{m=1}^{2M} (R_m / S_m) \qquad [6]$$

Equation [6] is computed over various values of n. H is estimated using the ordinary least squares (OLS) regression

$$\ln[(R/S)_n] = \ln(c) + H \ln(n) + \text{error} \qquad [7]$$

***Fluctuation analysis***

Estimation of the Hurst exponent using FA proceeds as follows. As before, subdivide the sample period T into M contiguous subperiods of n observations, and compute the following for m=1...M:

$$v_m = |p_{mn} - p_{m(n-1)}| \qquad [8]$$

where $\{p_t\}$ is the log price series. If Mn<T, compute a second set of M values of $\{v_m\}$ starting from the L+1th observation, indexed $\{v_{m+1}...v_{2M}\}$, where L is defined as before. The q'th-order partition function for time scale n is

$$S_q(T,n) = 2^{-1}\sum_{m=1}^{2M}(v_m)^q \qquad [9]$$

The FA focuses on the variation of $S_q(T,n)$ over changes in the time scale n, for several values of q. The scaling behaviour of $S_q(T,n)$ is investigated by examining the power law relationship

$$E[S_q(T,n)] = Tc(q)\times n^{Hq} = Tc(q)\times n^{\tau(q)+1} \qquad [10]$$

where c(q) is the prefactor and $\tau(q) = -1+Hq$ is the scaling function. The Hurst exponent is estimated using the fixed effects regression

$$\ln[S_q(T,n)] = a(q) + [-1+Hq] \ln(n) + \text{error} \qquad [11]$$

where $a(q) = \ln[nTc(q)]$.

***Other estimation methods***

Commonly used estimators of the order of fractional integration for a fractionally integrated time series were developed by Geweke and Porter-Hudak (henceforth GPH) (1983) and Robinson (1995). The properties of these estimators are compared by Andersson (2002), and several variants of the Robinson estimator are examined by Shimotsu and Phillips (2006).[9] These estimators are based on the periodogram as an estimator of the spectral density function of the returns series $\{x_t\}$. Let m denote the number of ordinates to be used in the estimation;[10] and let $I(\lambda_j) = (2\pi T)^{-1}\sum_{t=1}^{T} x_t \exp(-i\lambda_j t)$ denote the periodogram at the harmonic frequencies $\lambda_j = 2\pi j/T$ for j=1,...,m. The GPH estimator is the OLS estimator of d in the regression

$$\ln[I(\lambda_j)] = b_0 + d[-2\ln|1-\exp(-i\lambda_j)|] + \xi_j \qquad [12]$$

The Robinson estimator of d is

$$\ln[I(\lambda_j)] = a + b[\ln(\lambda_j)] + \nu_j \qquad [13]$$

where b = -2d. Kearns and Pagan (1997) identify the Pickands (1975), Hill (1975) and de Haan and Resnick (henceforth HR) (1980) estimators as the three most commonly used methods for the estimation of the characteristic exponent or tail index of an L-stable process.[11] Let $\{x_{(t)}\}$ denote the returns series reordered such that $x_{(1)} > x_{(2)} > ... > x_{(T)}$; and let m denote the number of observations in the upper tail to be used in the estimation.[12] The three estimators of H are

Pickands (1975): $[\ln(2)]^{-1}[\ln(x_{(m)} - x_{(2m)}) - \ln(x_{(2m)} - x_{(4m)})]$

Hill (1975): $\left[(m-1)^{-1}\sum_{i=1}^{m-1}\ln(x_{(i)})\right] - \ln(x_{(m)})$

HR (1980): $[\ln(m)]^{-1}[\ln(x_{(1)}) - \ln(x_{(m)})]$ [14]

The corresponding estimators of the characteristic exponent are obtained using the relation $\alpha = H^{-1}$.

## 3. Hypothesis tests for NIID returns against self-affine alternatives

Section 3 reports critical values for the statistical tests for departure from the NIID case based on estimation of the Hurst exponent using RRA and FA. Power functions for these tests are evaluated. Comparisons with alternative estimators for the order of fractional integration and for the tail index (see Section 2.4) are presented.

Hypothesis tests for a null hypothesis under which returns are NIID against an alternative under which returns are characterized either by [1] with d>0, or by [2] with $\alpha<2$, are based on the empirical distributions of the estimators of H obtained from [7] using RRA, or from [11] using FA. In both cases, the null and alternative hypotheses are $H_0$:H=0.5 and $H_1$:H>0.5. Critical values for these tests are obtained from the empirical distributions of these estimators, obtained from 5,000 replications of an NIID returns series. In all cases, the replications are generated for sample sizes T=1,000, 2,000, 5,000 and 10,000.[13]

Table 1 reports means, standard deviations and critical values for one-tail tests of $H_0$:H=0.5 against $H_1$:H>0.5 based on RRA and three alternative versions of FA. FA(1) is computed using q=(0.1,0.2, ...,1.0) in [9], [10] and [11]; FA(2) uses q=(0.3,0.6, ...,3.0); and FA(3) uses q=(0.5,1.0,...,5.0). The RRA produces upward-biased estimates of H. The magnitude of the bias decreases and the relative efficiency increases as T increases. FA(1), FA(2) and FA(3) produce downward-biased estimates of H. The magnitude of the bias is greatest for FA(1), followed by FA(2) and FA(3). In each case the magnitude of the bias decreases as T increases. The relative efficiency of the FA estimator is greatest for FA(3), followed by FA(2) and FA(1). In each case, the relative efficiency increases with T.

[insert Table 1 here]

Tables 2 and 3 report the mean values of the RRA and FA estimators of the Hurst exponent, when the true value of H exceeds 0.5. Each result is generated using 5,000 replications of a simulated self-affine returns series, based on [1] in the case where the process is ARFIMA(0,d,0), and based on [2] in the case where the process is L-stable with infinite higher-order moments. In each case, the replications are generated for H=0.54, 0.58 and 0.62.[14] For each estimation method, four sets of values for the mean estimated H are reported:

(i) ARFIMA(0,d,0) simulated returns based on [3];

(ii) ARFIMA(0,d,0) simulated returns based on [3] with a random re-ordering transformation applied, to preserve the distributional properties while eliminating long-range dependence;[15]

(iii) L-stable simulated returns based on [4];

(iv) L-stable simulated returns based on [4] with a normalizing transformation applied, to preserve long-range dependence while eliminating the non-Gaussian distributional properties.[16]

For (i), the mean estimated H are increasing with the true values of H, in a predictable manner. For RRA, the magnitude of the upward bias in the estimated H decreases somewhat as the true value of H increases; while for FA(1), FA(2) and FA(3) the magnitude of the downward bias is virtually unchanged as the true value of H increases.

For (ii), the mean estimated H for the randomly re-ordered ARFIMA series are virtually identical to the values reported in Table 1 for H=0.5 in the case of RRA, and slightly higher than the corresponding values in Table 1 for FA(1), FA(2) and FA(3). In each case, it is possible to interpret a discrepancy (similar to those shown in Tables 2 and 3) between the estimated H for an original data series and a randomly re-ordered transformation of the same series as evidence that returns are self-affine and characterized by long-range dependence.

For (iii), the mean estimated H obtained using FA(1) are increasing with the true values of H in a stable and predictable manner. The downward bias in the estimated H increases slightly as the true H increases. For RRA, FA(2) and FA(3), however, the mean estimated H *decreases* as the true value of H increases, indicating that these methods are unsuitable in the case where returns are L-stable with infinite higher-order moments. The source of the difficulty is that RRA, FA(2) and FA(3) examine the scaling behaviour of sample estimators of moments (q=2 only in the case of RRA, $\alpha<q\leq3$ for FA(2), and $\alpha<q\leq5$ for FA(3)) whose true values are infinite for an L-stable distribution with $\alpha<2$. This difficulty is avoided by FA(1), whose scope is restricted to $q<\alpha$ for all values of $\alpha$ that are considered in Table 2.

For (iv), the mean estimated H for the normalized L-stable series are virtually identical to the values reported in Table 1 for H=0.5 in the case of FA(1). It is possible to interpret a discrepancy (similar to those shown in Table 2) between the estimated H for an original returns series and a normalized transformation of the same series as evidence that returns are self-affine and L-stable with infinite higher-order moments.

[insert Tables 2 and 3 here]

The power functions of tests of $H_0$:H=0.5 against $H_1$:H>0.5 at the 0.05 significance level based on the RRA, FA(1), FA(2) and FA(3) estimators are examined in Table 4. For the case where the process under the alternative hypothesis is ARFIMA(0,d,0), RRA has superior power properties to FA(3). FA(3) is superior to FA(2), and FA(2) is superior to FA(1). For the case where the process under the alternative hypothesis is L-stable with infinite higher-order moments, however, only FA(1) has an appropriately shaped power function. The power functions for RRA, FA(2) and FA(3) tend rapidly towards zero as H increases, rendering these techniques unsuitable as a basis for testing for departure from NIID returns. Table 5 reports the power functions for the preferred estimator FA(1), at the 0.10, 0.05 and 0.01 significance levels.

[insert Tables 4 and 5 here]

Table 6 reports comparisons between the means and standard deviations of the GPH and Robinson estimators of d (see [12] and [13]) and the RRA, FA(1), FA(2) and FA(3) estimators of H, in the case where returns are ARFIMA(0,d,0). The GPH and Robinson estimators are both virtually unbiased, but Robinson is relatively more efficient than GPH. The upward-biased RRA estimator offers efficiency gains on both GPH and Robinson. The three downward-biased FA estimators are relatively less efficient than Robinson and RRA, but relatively more efficient than GPH.

[insert Table 6 here]

Table 7 reports comparisons between the means and standard deviations of the Pickands, Hill and HR estimators (see [14]) and the FA(1) estimator of the Hurst exponent, in the case where returns are L-stable with $\alpha \leq 2$, $H \geq 0.5$. Each of the Pickands, Hill and HR estimators is downward biased. Pickands is relatively inefficient over all values of H considered. HR is efficient for H=0.5, but is relatively inefficient for H>0.5. Hill is relatively efficient over all values of H considered, and offers a modest efficiency gain over FA(1). The latter is also downward biased, but to a lesser degree than the other three estimators. Although the Hill estimator is preferred to FA(1) on the criterion of relative efficiency, the Hill estimator is a less reliable estimator of H than FA(1) in the case where the probability distribution for returns is independent, but non-Gaussian with finite higher-order moments. In this case, the true value of H is 0.5. The downward bias in the Hill estimator is diminished, creating a tendency to reject $H_0$:H=0.5 falsely in favour of $H_1$:H>0.5. In contrast, the downward bias in the FA(1) estimator is virtually unaffected. In Monte Carlo simulations for returns drawn from the student t-distribution with either 10 or 20 degrees of freedom and T=5,000, the rejection rates for the test of $H_0$:H=0.5 in favour of $H_1$:H>0.5 based on the Hill estimator, using a significance level of 0.05 and critical values based on simulated NIID returns, were 0.978 and 0.599 respectively. The rejection rates for the test based on the FA(1) estimator were 0.05 in both cases.

As an estimator of the order of fractional integration, the FA(1) estimator is less reliable than both Robinson and RRA, but more reliable than GPH. As an estimator of the characteristic exponent or tail index of an L-stable process, the FA(1) estimator is less reliable than Hill if returns are either NIID or L-stable, but more reliable than either Pickands or HR. FA(1) is considerably more reliable than Hill if returns are independent but non-Gaussian with finite higher-order moments. As an estimator of the Hurst exponent, FA(1) is unique (among the estimators considered in this section) in terms of its reliability under both of the fractionally integrated and L-stable alternatives to the NIID null hypothesis.

[insert Table 7 here]

## 4. Estimation of the Hurst exponent for 11 stock market indices

Section 4 reports an application of the techniques described in this paper, using daily logarithmic returns data calculated from closing prices for 11 developed country stock market indices for the period July 1987 to May 2011 (inclusive). For convenience we assign the 11 stock markets to two categories by market capitalization. The three large-capitalization markets are Japan (represented by the Nikkei index), the UK (FTSE 100) and the US (SP500); and the eight small-capitalization markets are France (CAC), Finland (OMX Helsinki 25), Germany (DAX), Ireland (ISEQ), Italy (MIBTel), Netherlands (AMX), Spain (Madrid SE General), and Sweden (OMX Stockholm 30).[17] In a cross-country analysis, Cajueiro and Tabak (2004, 2005) interpret estimated Hurst exponents for either stock returns or volatility as indicators of stock market efficiency. We posit an association between market size and market efficiency, such that returns for the large-capitalization markets exhibit the least evidence, and those for the small-capitalization markets exhibit the strongest evidence of long-range dependence.

Table 8 reports the sample means and standard deviations, and sample skewness and kurtosis coefficients, for the daily logarithmic returns series. Table 9 reports the Hurst exponent estimates for the returns series on the 11 stock market indices obtained using the RRA, FA(1), FA(2) and FA(3) estimators. For comparison purposes, the Robinson (1995) estimator of d (see [13]) and the Hill (1975) estimator of H (see [14]) are also reported.

It is well known that the identification of long-range dependence in the presence of short-range dependence is challenging, owing to difficulties in disentangling the short-range and long-range dependence components (Smith et al., 1997). In some previous studies, estimators of H are applied to the residuals of a fitted autoregressive model for the returns series, to eliminate short-range dependence by filtering before testing for long-range dependence (Jacobsen, 1996; Opong *et al.*, 1999). In the present study we apply the long-range dependence estimators to the returns series both with and without filtering. We compare the estimated H for filtered returns with critical values constructed using NIID Monte Carlo simulations; and we compare the estimated H for unfiltered returns with critical values constructed using recursive Monte Carlo simulations, in which the simulated series have a short-range dependence structure that corresponds to a fitted autoregressive model for the actual returns series for each index.

Since filtering tends to eliminate a portion of the long-range dependence when the latter is present, an estimated H exponent that is significantly different from H=0.5 (for a pre-filtered returns series using NIID critical values) should constitute strong evidence of long-range dependence. Estimation of H using an unfiltered returns series leaves open the possibility of conflating short-range and long-range dependence. Critical values based on simulated series with a short-range dependence structure imposed, based on the coefficients obtained by fitting a (short-range) autoregressive model to the original series, will tend to be inflated, because the estimated short-range autoregressive coefficients are overstated if long-range dependence is present. Accordingly, an estimated H for an unfiltered series that is significantly different from H=0.5 when compared with critical values derived from simulated series with short-range dependence imposed should also constitute strong evidence of long-range dependence.

In view of the results reported in Section 3 of this paper, the FA(1) estimator is considered the best equipped to distinguish between the cases H>0.5 and H=0.5, if the process in the case H>0.5 is unknown and could be either ARFIMA(0,d,0) or L-stable with infinite higher-order moments. In the ARFIMA(0,d,0) case, however, the RRA, FA(2) and FA(3) estimators are more powerful than the FA(1) estimator. Therefore all four sets of Hurst exponent estimates are of interest, and all four sets are reported in Table 9.

Using a significance level of 0.05, the FA(1) Hurst exponent estimates based on unfiltered returns support the rejection of $H_0$:H=0.5 in favour of $H_1$:H>0.5 for one of the three large-capitalization markets, and for five of the eight small-capitalization markets. The corresponding estimates based on filtered returns support the rejection of the same null for none of the large-capitalization markets, and for four of the small-capitalization markets. On the basis of rejection of this null in the tests based on both sets of FA(1) estimates, we infer that there is strong evidence of self-affine scaling behaviour for Finland, Germany, Ireland and Sweden. On the basis of rejection in the tests based on FA(1) estimates using unfiltered returns only, we infer that there is weak evidence of self-affine scaling behaviour for the US and the Netherlands.

For Finland, Germany, Ireland and Sweden, the tests based on the RRA estimator using both unfiltered and filtered returns also reject the null hypothesis of H=0.5 in every case. The tests based on the FA(2) estimator using unfiltered returns reject this null for Finland, Ireland and Sweden, and the tests based on the FA(2) estimator using filtered returns reject for Finland and Ireland. The tests based on the FA(3) estimator fail to reject, however, in every case. These patterns suggest that the evidence of self-affine scaling behaviour might be attributable to long-range dependence, rather than with returns having been drawn from an L-stable distribution with infinite higher-order moments. In the latter case we should expect all of the tests based on the RRA, FA(2) and FA(3) estimators to fail to reject the null hypothesis of H=0.5.[18]

For the US and the Netherlands, the tests based on the RRA, FA(2) and FA(3) estimators using unfiltered returns fail to reject the null hypothesis of H=0.5. These patterns suggest that the finding of self-affine scaling behaviour in the test based on the FA(1) estimator might be attributable to returns having been drawn from an L-stable distribution with infinite higher-order moments, rather than long-range dependence. For the US in particular, this interpretation seems consistent with an extremely large sample kurtosis coefficient reported in Table 8. The evidence that there is long-range dependence for four of the eight small-capitalization markets, and none of the three large-capitalization markets, seems consistent with the posited link between market size and market efficiency.

Finally, the Robinson estimator fails to reject the null hypothesis $H_0$:d=0 in favour of $H_1$:d>0 for any of the estimations. The Hill estimator rejects the null hypothesis $H_0$:H=0 in favour of $H_1$:H>0 consistently throughout the entire sample period. In view of the evidence that Hill is unreliable in distinguishing between different forms of non-Gaussian behaviour, however, it is possible to infer from the results from the Hill estimator only that returns are non-Gaussian, but not that returns are L-stable.

[insert Table 9 here]

## 5. Conclusion

This paper develops hypothesis tests for departures from null hypothesis of NIID logarithmic returns for the case where returns are self-affine under the alternative hypothesis. In this case the distributions of returns measured over different time scales (daily, monthly, yearly, and so on) are identical, except for a single non-random contraction that depends on the time scale only. The scaling properties of a returns series are conveniently summarized by the Hurst exponent. A self-affine returns series might be either fractionally integrated, in which case returns exhibit long-range dependence; or L-stable, in which case returns are characterized by random drawings from a distribution with infinite variance and higher-order moments.

Tests for the null hypothesis of NIID returns against alternatives in which returns are self-affine are based on the application of two methods for the identification of scaling behaviour that have been used widely in the previous literature: rescaled range analysis (RRA), and fluctuation analysis (FA). Previously, researchers have reported evidence in the form of point estimates of the Hurst exponent, or graphical analysis of returns data, without having any basis for the evaluation of the statistical significance of departures from the NIID case. This paper addresses this deficiency in the empirical literature.

The principal findings are as follows. The performance of tests for departure from the NIID case based on RRA and FA is satisfactory when returns are self-affine and characterized by long-range dependence. In this case, the test based on RRA has greater power than tests based on FA. However, the test based on RRA performs poorly when returns are self-affine and characterized as L-stable with infinite higher-order moments. In this case, the choice of sample moments over which the FA is computed is crucial: the FA should not consider moments whose true values are infinite. The use of RRA is inappropriate in this case because RRA is based on an examination of the sample scaling behaviour of the second moment, whose true value is infinite. As an estimator of the Hurst exponent, the variant of the FA that considers finite sample moments only is uniquely reliable (among the estimators considered in this paper) under both of the fractionally integrated and L-stable alternatives to the NIID null hypothesis. These finite-sample properties of the estimators when returns exhibit either form of self-affinity can be exploited to determine the source of self-affinity in empirical returns data.

The techniques are illustrated by means of an analysis of the fractal properties of the daily logarithmic returns for the indices of 11 stock markets, three of which are classified as large in terms of market capitalization, and eight as small. We find strong evidence of self-affine scaling behaviour for four markets, Finland, Germany, Ireland and Sweden. In all four cases, long-range dependence appears to be the source of the self-affine scaling behaviour. We find weak evidence of self-affine scaling behaviour in two further cases, the US and the Netherlands, for which the results are consistent with returns having been drawn from an L-stable distribution with infinite higher-order moments, rather than long-range dependence.

## Notes

[1] This literature reports empirical evidence on the fractal properties of stock market and individual company returns (Barkoulas and Baum, 1996; Di Matteo, Aste, and Dacorogna, 2005), commodity prices (Alvarez-Ramirez *et al.*, 2002), inflation rates (Lee, 2005), and currency exchange rates (Fisher, Calvet, and Mandelbrot, 1997; Batten and Ellis, 2001; Calvet and Fisher 2002).

[2] FBM is a generalization of Brownian Motion, the continuous-time analogue of the random walk. FBM has increments that are long-range dependent and therefore non-random (Mandelbrot and van Ness, 1968).

[3] RRA was introduced by Hurst (1951). Refinements are suggested by Mandelbrot and Wallis (1968, 1969a,b,c), Mandelbrot (1972, 1975), Mandelbrot and Taqqu (1979), and Lo (1991).

[4] This study uses the variant of FA employed by Mandelbrot, Fisher and Calvet (1997). Recent methodological contributions for the estimation of the long-range dependence parameter using FA include Fillol and Tripier (2004) and Fillol (2007).

[5] See Greene and Fielitz (1977), Peters (1991), McKenzie (2001), Alvarez-Ramirez *et al.* (2002), Costa and Vasconcelos (2003), Kim and Yoon (2004) and Norouzzadeh and Jafari (2005).

[6] The Cauchy distribution has ($\alpha$=1, $\beta$=0); and the Levy distribution, also known as the Pareto-Levy distribution, has ($\alpha$=0.5, $\beta$=1) or ($\alpha$=0.5, $\beta$=–1).

[7] Recent applications of Monte Carlo techniques in analysing long-range dependence or L-stable processes include Baillie and Kapetanios (2007), Ndongo *et al.* (2009), Dufour and Kurz-Kim (2010), Barounik and Kristoufek (2010) and Iacone (2010).

[8] If Mn = N, $R_{m+M} = R_m$ and $S_{m+M} = S_m$ for m = 1,...,M.

[9] See also Beran (1992), Cheung and Diebold (1994), Crato and Ray (1996), Dalhaus (1989), Fox and Taqqu (1986), Hiemstra and Jones (1997), Richards (2000) and Sowell (1992).

[10] All results reported in this paper are based on $m=T^{0.5}$ for the GPH estimator, and $m=T^{0.9}$ for the Robinson estimator.

[11] See also Dekkers and de Haan (1989), DuMouchel (1983), Hols and De Vries (1991), Hsu, Miller and Wichern (1974) and Pagan (1996).

[12] All results for the Pickands, Hill and HR estimators reported in this paper are based on m=0.05T.

[13] The time scales for which [6] and [9] are computed are such that ln(n) increases from $n_{MIN}$ to $n_{MAX}$ in steps of 0.15, where $\ln(n_{MIN})=1.6$ and $\ln(n_{MAX})=0.15\text{int}[\{\ln(0.1T)\}/0.15]$, where int[ ] is the next-lowest integer.

[14] These values correspond to d=0.04, 0.08 and 0.12 respectively in [1], and $\alpha$=1.85, 1.72 and 1.61 in [2].

[15] Using a random-number generator, create $\xi_t \sim U(0,1)$ for t=1,...,T. Let r(t) denote the rank of $\xi_t$ among $\{\xi_1,\ldots, \xi_t\}$. The randomly re-ordered transformation of the original returns series $\{y_t\}$ is $y_t^* = y_{r(t)}$.

[16] Let $\rho(t)$ denote the rank of $y_t$ among $\{y_1,\ldots,y_T\}$. The normalized transformation of $\{y_t\}$ is $y_t^{**} = \Phi^{-1}(\rho(t)/(T+1))$, where $\Phi^{-1}(\ )$ is the inverse of the standard normal distribution function.

[17] The data for the closing daily prices of the stock market indices are obtained from Thomson One Banker. End-of-year market capitalization data (in USD million) for the associated stock markets are as follows. US: 13,394,082 (NYSE-Euronext US, 2010); Japan: 3,827,774 (2010); the UK: 1,868,153 (2008); France: 1,489,520 (2008); Germany: 1,429,719 (2010); Italy: 655,848 (2009); Spain: 1,171,625 (2010); Finland: 118,167 (2003); Ireland: 60,368 (2010); Netherlands: 393,238 (2008); Sweden: 170,283 (2003). Data sources: World Stock Exchanges website (http://www.world-exchanges.org/statistics/time-series/market-capitalization) and stock exchanges websites.

[18] Informal comparisons (not reported in Table 9) between the FA(1) Hurst exponent estimates, and the FA(1) estimates obtained from randomly re-ordered and normalized transformations of the original returns series, support this interpretation. In most cases there are large differences between the estimated H for the original series and for a randomly re-ordered transformation; and small differences between the estimated H for the original series and for a normalized transformation.

## References

Alvarez-Ramirez, Jose, Myriam Cisneros, Carlos Ibarra-Valdez, and Angel Soriano, "Multifractal Hurst analysis of crude oil prices," *Physica A* 313 (2002), 651-670.

Andersson, Jonas, "An improvement of the GPH estimator," *Economics Letters* 77 (2002), 137-146.

Baillie, Richard T., and George Kapetanios, "Testing for neglected nonlinearity in long-memory models," *Journal of Business and Economic Statistics* 25 (2007), 447-461.

Banerjee, Anindya, and Giovanni Urga, "Modelling structural breaks, long memory and stock market volatility: an overview," *Journal of Econometrics* 129 (2005), 1-34.

Barkoulas, John T., and Christopher F. Baum, "Long-term dependence in stock returns," *Economics Letters* 53 (1996), 253-259.

Barnard, G. A., "Comment on 'the spectral analysis of point processes' by M.S. Bartlett," *Journal of the Royal Statistical Society B* 25 (1963), 294.

Barunik, Jozef, and Ladislav Kristoufek, "On Hurst exponent estimation under heavy-tailed distributions," *Physica A* 389 (2010), 3844-3855.

Batten, Jonathan A., and Craig Ellis, "Scaling relationships of Gaussian processes," *Economics Letters* 72 (2001), 291-296.

Beran, Jan, "Statistical methods for data with long range dependence," *Statistical Science* 7 (1992), 404-427.

Birnbaum, Z. H., "Computers and unconventional test-statistics," In Proschan, F., and R.J Serfling (Eds.), *Reliability and Biometry*. SIAM, Philadelphia, PA (1974), 441–458.

Black, Fisher, and Myron Scholes, "The valuation of option contracts and a test of market efficiency," *Journal of Finance*, 27 (1972), 399-418.

Black, Fisher, and Myron Scholes, "The pricing of options and corporate liabilities," *Journal of Political Economy*, 81 (1973), 418-637.

Cajueiro, Daniel O., and Benjamin M. Tabak, "Ranking efficiency for emerging markets," *Chaos, Solitons and Fractals* 22 (2004), 349-352.

Cajueiro, Daniel O., and Benjamin M. Tabak, "Ranking efficiency for emerging equity markets II," Chaos, Solitons and Fractals 23 (2004), 671-675.

Calvet, Laurent, and Adlai Fisher, "Multifractality in asset returns: theory and evidence," *The Review of Economics and Statistics* 84 (2002), 381-406.

Chambers, J. M., Mallows, C.L., and B. W. Stuck, "A method for simulating stable random variables," *Journal of American Statistical Association* 71 (1976), 340-344.

Cheung, Yin-Wong, and Francis X. Diebold, "On maximum likelihood estimation of the differencing parameter of fractionally-integrated noise with unknown mean," *Journal of Econometrics* 62 (1994), 301-316.

Cont, Rama, and Peter Tankov (2004). *Financial modelling with jump processes*. London. Chapman Hall.

Costa, Rogerio L., and G.L. Vasconcelos, "Long-term correlations and nonstationarity in the Brazilian stock market," *Physica A* 329 (2003), 231-248.

Crato, Nuno, and Bonnie K. Ray, "Model selection and forecasting for long-range dependent processes," *Journal of Forecasting* 15 (1996), 107-125.

Dahlhaus, Rainer, "Efficient parameter estimation for self-similar processes," *The Annals of Statistics* 17 (1989), 1749-1766.

de Haan L., and S.I. Resnick, "A Simple Asymptotic Estimate for the Index of a Stable Distribution," *Journal of the Royal Statistical Society,* series B (1980) 42, 83-87.

Dekkers, Arnold L.M., and Laurens de Haan, "On the estimation of the extreme-value index and large quantile estimation," *The Annals of Statistics* 17 (1989), 1795-1832.

Di Matteo, T., Aste, T., and M. M. Dacorogna, "Long-range memories of developed and emerging markets: using the scaling analysis to characterize their stage of development," *Journal of Banking and Finance* 29 (2005), 827-851.

Dufour, Jean-Marie, "Monte Carlo tests with nuisance parameters: a general approach to finite-sample inference and standard asymptotics in econometrics," *Journal of Econometrics* 133 (2006), 443-477.

Dufour, Jean-Marie, and Jeong-Ryeol Kurz-Kim, "Exact inference and optimal invariant estimation for the stability parameter of symmetric α-stable distributions," *Journal of Empirical Finance* 17 (2010), 180-194.

DuMouchel, William H., "Estimating the stable index $\alpha$ in order to measure tail thickness: a critique," *The Annals of Statistics* 11 (1983), 1019-1031.

Dwass, Meyer, "Modified randomization tests for nonparametric hypotheses," *The Annals of Mathematical Statistics* 28 (1957), 181-187.

Fama, Eugene F., "Efficient capital markets: a review of theory and empirical work," *Journal of Finance* 25 (1970), 383-417.

Fillol, Jerome, and Fabien Tripier, "The scaling function-based estimator of the long memory parameter: a comparative study," *Economics Letters* 84 (2004), 49-54.

Fillol, Jerome, "Estimating long memory: Scaling function vs Andrews and Guggenberger GPH," *Economics Letters* 95 (2007), 309-314.

Fisher, Adlai, Laurent Calvet, and Benoit Mandelbrot, "Multifractality of Deutschemark / US dollar exchange rates," Yale University, Cowles Foundation Discussion Paper #1166 (1997).

Fox, Robert, and Murad S. Taqqu, "Large-sample properties of parameter estimates for strongly dependent stationary Gaussian time series," *The Annals of Statistics* 14 (1986), 517-532.

Geweke, John, and Susan Porter-Hudak, "The estimation and application of long memory models and fractional integration," *Journal of Time series Analysis* 4 (1983), 221-238.

Granger, Clive W. J., and Roselyne Joyeux, "An introduction to long memory time series models and fractional differencing," *Journal of Time Series Analysis* 1 (1980), 15-29.

Greene, Myron T., and Bruce D. Fielitz, "Long-range autocorrelation in common stock returns," *Journal of Financial Economics* 4 (1977), 339-349.

Hiemstra, Craig, and Jonathan D. Jones, "Another look at long memory in common stock returns," *Journal of Empirical Finance* 4 (1997), 373-401.

Hill, Bruce M., "A simple general approach to inference about the tail of a distribution," *The Annals of Statistics* 3 (1975), 1163-1174.

Hols, Martien C.A.B., and Casper G. De Vries, "The limiting distribution of extremal exchange rate returns," *Journal of Applied Econometrics* 6 (1991), 287-302.

Hope, Adery C. A., "A simplified Monte Carlo significance test procedure," *Journal of the Royal Statistical Society B* 30 (1968), 582-598.

Hosking, Jonathan R.M., "Fractional differencing," *Biometrika* 68 (1981), 165-176.

Hsu, Der-Ann, Robert B. Miller, and Dean W. Wichern, "On the stable Paretian behavior of stock market prices," *Journal of the American Statistical Association* 69 (1974), 108-113.

Hurst, H.E., "The long-term storage capacity of reservoirs," *Transaction of the American Society of Civil Engineer* (1951).

Iacone, Fabrizio, "Local Whittle estimation of the memory parameter in presence of deterministic components," *Journal of Time Series Analysis* 31 (2010), 37-49.

Jacobsen, Ben, "Long term dependence in stock returns," *Journal of Empirical Finance* 3 (1996), 393-417.

Kearns, Phillip, and Adrian Pagan, "Estimating the density tail index for financial time series," *The Review of Economics and Statistics* 79 (1997), 171-175.

Kim, Kyungsik, and Seong-Min Yoon, "Multifractal features of financial markets," *Physica A* 344 (2004), 272-278.

Levy, Paul (1925), Calcul des probabilities, Gauthier-Villars, Paris.

Lee, Jin, "Estimating memory parameter in the US inflation rate," *Economics Letters* 87 (2005), 207-210.

Lintner, John, "The valuation of risk assets and the selection of risky investments in stock portfolios and capital budgets," *Review of Economics and Statistics* 47 (1965), 13-37.

Lo, Andrew W., "Long-term memory in stock market prices," *Econometrica* 59 (1991), 1279-1313.

Mandelbrot, Benoit, "The variation of certain speculative prices," *Journal of Business* 36 (1963), 394-413.

Mandelbrot, Benoit, "The variation of some other speculative prices," *Journal of Business* 40 (1967), 393-413.

Mandelbrot, Benoit, "When can price be arbitraged efficiently? A limit to the validity of the random walk and martingale models," *The Review of Economics and Statistics* 53 (1971), 225-236.

Mandelbrot, Benoit, "Statistical methodology for non-periodic cycles: from the covariance to R/S analysis," *Annals of Economic and Social Measurement* 1 (1972), 259-290.

Mandelbrot, Benoit, "Limit theorems on the self-normalized range for weakly and strongly dependent processes," *Zeitschrift fur Wahrscheinlichkeitstheorie und Verwandte Gebiete* (1975), 271-285.

Mandelbrot, Benoit (1977). *Fractals, form, chance and dimension*. New York: W.H. Freeman.

Mandelbrot, Benoit, Adlai Fisher, and Laurent Calvet, "A multifractal model of asset returns," Yale University (1997), Cowles Foundation Discussion Paper #1164.

Mandelbrot, Benoit, and Murad S. Taqqu, "Robust R/S analysis of long-run serial correlation," *Bulletin of the International Statistical Institute* 48 (1979), Book 2, 59-104.

Mandelbrot, Benoit, and John W. van Ness, "Fractional Brownian Motion, Fractional Noises and application," *SIAM Review* 10 (1968), 422-437.

Mandelbrot, Benoit, and James R. Wallis, "Noah, Joseph and operational hydrology," *Water Resources Research* 4 (1968), 909-918.

Mandelbrot, Benoit, and James R. Wallis, "Computer experiments with Fractional Gaussian Noises. Parts 1,2,3," *Water Resources Research* 5 (1969a), 228-267.

Mandelbrot, Benoit, and James R. Wallis, "Some long run properties of geophysical records," *Water Resources Research* 5 (1969b), 321-340.

Mandelbrot, Benoit, and James R. Wallis, "Robustness of the rescaled range R/S in the measurement of noncyclic long run statistical dependence," *Water Resources Research* 5 (1969c), 967-988.

McKenzie, Michael D., "Non-periodic Australian stock market cycles: evidence from rescaled range analysis," *Economic Record* 77 (2001), 393-406.

Ndongo, Mor, Abdou Ka Diongue, Diop Adiou, and Simplice Dossou-Gbete, "Estimation of long-memory parameters for seasonal fractional ARIMA with stable innovations," *Statistical Methodology* 7 (2010), 141-151.

Norouzzadeh, P., and G. R. Jafari, "Application of multifractal measures to Tehran price index," *Physica A* 356 (2005), 609-627.

Opong, Kuaku K., Mulholland, Gwyneth, Fox, Alan F., and Kambiz Farahmand, "The behaviour of some UK equity indices: an application of Hurst and BDS tests," *Journal of Empirical Finance* 6 (1999), 267-282.

Pagan, Adrian, "The econometrics of financial markets," *Journal of Empirical Finance* 3 (1996), 15-102.

Peters, Edgar E. (1991). *Chaos and order in the capital markets*. New York. John Wiley & Sons.

Peters, Edgar E. (1994). *Fractal markets analysis: applying chaos theory to investments and economics*. New York. John Wiley & Sons.

Pickands, James, "Statistical inference using extreme order statistics," *The Annals of Statistics* 3 (1975), 119-131.

Richards, Gordon R., "The fractal structure of exchange rates: measurement and forecasting," *Journal of International Financial Markets, Institutions and Money* 10 (2000), 163–180.

Robinson, Peter M., "Gaussian semiparametric estimation of long range dependence," *The Annals of Statistics* 23 (1995), 1630-1661.

Rogers, L.C.G., "Arbitrage with Fractional Brownian Motion," *Mathematical Finance* 7 (1997), 95-105.

Sharpe, William F., "Capital asset prices: a theory of market equilibrium under conditions of risk," *Journal of Finance* 19 (1964), 425-442.

Shimotsu, Katsumi, and Peters C.B. Phillips, "Local Whittle estimation of fractional integration and some of its variants," *Journal of Econometrics* 130 (2006), 209-233.

Sowell, Fallaw B., "Maximum likelihood estimation of stationary univariate fractionally-integrated time-series models," *Journal of Econometrics* 53 (1992), 165-188.

Weron, Rafal, "On the Chambers-Mallows-Stuck method for simulating skewed stable random variables," *Statistics and Probability Letters* 28 (1996), 165-171.

Table 1 Means, standard deviations and critical values (90th, 95th and 99th percentiles) for estimated Hurst exponents under the null hypothesis of NIID returns (H=0.5)

| Sample, T | Mean | Standard deviation | Critical values for significance levels: 0.10 | 0.05 | 0.01 |
|---|---|---|---|---|---|
| Rescaled range analysis, RRA | | | | | |
| 1000 | .613 | .020 | .639 | .646 | .659 |
| 2000 | .595 | .016 | .616 | .621 | .632 |
| 5000 | .578 | .012 | .594 | .598 | .606 |
| 10000 | .568 | .010 | .580 | .584 | .590 |
| Fluctuation analysis, q=(0.1,0.2,...,1.0), FA(1) | | | | | |
| 1000 | .454 | .062 | .532 | .555 | .594 |
| 2000 | .477 | .050 | .540 | .557 | .588 |
| 5000 | .480 | .037 | .528 | .541 | .562 |
| 10000 | .481 | .031 | .522 | .533 | .554 |
| Fluctuation analysis, q=(0.3,0.6,...,3.0), FA(2) | | | | | |
| 1000 | .480 | .058 | .552 | .572 | .608 |
| 2000 | .488 | .048 | .548 | .565 | .594 |
| 5000 | .491 | .036 | .536 | .548 | .570 |
| 10000 | .491 | .030 | .530 | .541 | .557 |
| Fluctuation analysis, q=(0.5,1.0,...,5.0), FA(3) | | | | | |
| 1000 | .476 | .060 | .552 | .572 | .613 |
| 2000 | .483 | .049 | .546 | .563 | .591 |
| 5000 | .488 | .037 | .535 | .548 | .573 |
| 10000 | .489 | .031 | .529 | .540 | .558 |

Notes

Each result is based on 5,000 replications of a simulated NIID returns series. The RRA is described by equations [5] to [7]. The FA is described by [8] to [11].

Table 2 Mean estimated Hurst exponents, RRA and FA(1)

| | Rescaled range analysis, RRA | | | | Fluctuation analysis, q=(.1,.2,...,1.0), FA(1) | | | |
|---|---|---|---|---|---|---|---|---|
| Sample, T | ARFIMA | ARFIMA re-order | L-stable | L-stable normalize | ARFIMA | ARFIMA re-order | L-stable | L-stable normalize |
| H=0.54 | | | | | | | | |
| 1000 | .634 | .612 | .607 | .613 | .494 | .461 | .493 | .456 |
| 2000 | .619 | .595 | .589 | .595 | .516 | .484 | .513 | .476 |
| 5000 | .605 | .578 | .573 | .578 | .519 | .486 | .517 | .480 |
| 10000 | .596 | .568 | .563 | .568 | .521 | .487 | .519 | .480 |
| H=0.58 | | | | | | | | |
| 1000 | .656 | .612 | .602 | .613 | .534 | .473 | .530 | .456 |
| 2000 | .643 | .595 | .584 | .595 | .556 | .495 | .550 | .476 |
| 5000 | .631 | .578 | .569 | .578 | .559 | .496 | .554 | .480 |
| 10000 | .624 | .568 | .559 | .568 | .560 | .496 | .556 | .480 |
| H=0.62 | | | | | | | | |
| 1000 | .678 | .612 | .597 | .613 | .574 | .491 | .565 | .456 |
| 2000 | .667 | .595 | .580 | .595 | .596 | .513 | .586 | .476 |
| 5000 | .658 | .578 | .565 | .578 | .599 | .513 | .591 | .480 |
| 10000 | .652 | .568 | .555 | .568 | .600 | .511 | .594 | .480 |

Notes

Each result is based on 5,000 replications of a simulated self-affine returns series. The RRA is described by equations [5] to [7]. The FA is described by [8] to [11].

Table 3 Mean estimated Hurst exponents, FA(2) and FA(3)

| | Fluctuation analysis, q=(.1,.2,...,1.0), FA(2) | | | | Fluctuation analysis, q=(.1,.2,...,1.0), FA(3) | | | |
|---|---|---|---|---|---|---|---|---|
| Sample, T | ARFIMA | ARFIMA re-order | L-stable | L-stable normalize | ARFIMA | ARFIMA re-order | L-stable | L-stable normalize |
| H=0.54 | | | | | | | | |
| 1000 | .519 | .487 | .464 | .482 | .515 | .483 | .399 | .479 |
| 2000 | .527 | .495 | .467 | .488 | .522 | .491 | .392 | .483 |
| 5000 | .530 | .496 | .466 | .491 | .527 | .493 | .379 | .488 |
| 10000 | .531 | .496 | .462 | .490 | .528 | .494 | .368 | .488 |
| H=0.58 | | | | | | | | |
| 1000 | .558 | .498 | .454 | .482 | .554 | .494 | .358 | .479 |
| 2000 | .567 | .506 | .456 | .488 | .561 | .501 | .351 | .483 |
| 5000 | .569 | .506 | .454 | .491 | .566 | .502 | .339 | .488 |
| 10000 | .570 | .505 | .451 | .490 | .567 | .502 | .330 | .488 |
| H=0.62 | | | | | | | | |
| 1000 | .598 | .513 | .447 | .482 | .593 | .508 | .334 | .479 |
| 2000 | .606 | .521 | .450 | .488 | .600 | .515 | .329 | .483 |
| 5000 | .609 | .520 | .448 | .491 | .605 | .516 | .319 | .488 |
| 10000 | .609 | .518 | .445 | .490 | .606 | .514 | .313 | .488 |

Notes

See notes to Table 2.

Table 4 Power of one-tail tests of $H_0$:H=0.5 against $H_1$:H>0.5, significance level = 0.05, all estimation methods

| | ARFIMA under alternative hypothesis | | | | L-stable under alternative hypothesis | | | |
|---|---|---|---|---|---|---|---|---|
| Sample, T | RRA | FA(1) | FA(2) | FA(3) | RRA | FA(1) | FA(2) | FA(3) |
| H=0.54 | | | | | | | | |
| 1000 | .290 | .162 | .186 | .172 | .025 | .180 | .032 | .013 |
| 2000 | .448 | .214 | .225 | .214 | .020 | .205 | .021 | .006 |
| 5000 | .717 | .297 | .312 | .295 | .017 | .280 | .014 | .001 |
| 10000 | .876 | .348 | .373 | .362 | .014 | .331 | .010 | .000 |
| H=0.58 | | | | | | | | |
| 1000 | .685 | .378 | .425 | .392 | .010 | .357 | .020 | .004 |
| 2000 | .912 | .496 | .523 | .496 | .007 | .423 | .011 | .001 |
| 5000 | .996 | .675 | .707 | .678 | .004 | .592 | .004 | .000 |
| 10000 | 1.000 | .790 | .815 | .797 | .002 | .687 | .003 | .000 |
| H=0.62 | | | | | | | | |
| 1000 | .931 | .617 | .672 | .641 | .005 | .538 | .015 | .001 |
| 2000 | .996 | .762 | .787 | .757 | .003 | .644 | .009 | .000 |
| 5000 | 1.000 | .918 | .936 | .922 | .001 | .819 | .002 | .000 |
| 10000 | 1.000 | .968 | .976 | .972 | .000 | .899 | .001 | .000 |

Notes

See notes to Table 2.

Table 5 Power of one-tail tests of $H_0$:H=0.5 against $H_1$:H>0.5, various significance levels, Fluctuation analysis, q=(0.1,0.2,...,1.0), FA(1)

| | ARFIMA under alternative hypothesis | | | L-stable under alternative hypothesis | | |
|---|---|---|---|---|---|---|
| | Significance level: | | | Significance level: | | |
| Sample, T | 0.10 | 0.05 | 0.01 | 0.10 | 0.05 | 0.01 |
| H=0.54 | | | | | | |
| 1000 | .274 | .162 | .053 | .275 | .180 | .071 |
| 2000 | .328 | .214 | .080 | .303 | .205 | .092 |
| 5000 | .414 | .297 | .138 | .378 | .280 | .138 |
| 10000 | .499 | .348 | .159 | .451 | .331 | .165 |
| H=0.58 | | | | | | |
| 1000 | .525 | .378 | .170 | .482 | .357 | .189 |
| 2000 | .630 | .496 | .280 | .545 | .423 | .250 |
| 5000 | .775 | .675 | .476 | .685 | .592 | .413 |
| 10000 | .871 | .790 | .588 | .779 | .687 | .494 |
| H=0.62 | | | | | | |
| 1000 | .739 | .617 | .385 | .647 | .538 | .345 |
| 2000 | .845 | .762 | .565 | .740 | .644 | .454 |
| 5000 | .956 | .918 | .810 | .880 | .819 | .679 |
| 10000 | .985 | .968 | .902 | .941 | .899 | .778 |

Notes

See notes to Table 2.

Table 6 Mean and standard deviation of various estimators of the order of fractional integration or Hurst exponent, fractionally integrated process

| Estimator | Mean | | | | Standard deviation | | | |
|---|---|---|---|---|---|---|---|---|
| Parameter values | d=0.00 H=0.50 | d=0.04 H=0.54 | d=0.08 H=0.58 | d=0.12 H=0.62 | d=0.00 H=0.50 | d=0.04 H=0.54 | d=0.08 H=0.58 | d=0.12 H=0.62 |
| Sample, T=2000 | | | | | | | | |
| GPH, d | -.000 | .040 | .080 | .121 | .111 | .112 | .112 | .112 |
| Robinson, d | .000 | .037 | .073 | .110 | .022 | .022 | .022 | .022 |
| RRA, H | .595 | .619 | .643 | .667 | .016 | .016 | .017 | .017 |
| FA(1), H | .477 | .516 | .556 | .595 | .050 | .051 | .053 | .055 |
| FA(2), H | .488 | .527 | .567 | .606 | .048 | .049 | .050 | .051 |
| FA(3), H | .483 | .522 | .561 | .600 | .049 | .050 | .051 | .053 |
| Parameter values | d=0.00 H=0.50 | d=0.04 H=0.54 | d=0.08 H=0.58 | d=0.12 H=0.62 | d=0.00 H=0.50 | d=0.04 H=0.54 | d=0.08 H=0.58 | d=0.12 H=0.62 |
| Sample, T=5000 | | | | | | | | |
| GPH, d | .000 | .040 | .081 | .122 | .085 | .085 | .085 | .085 |
| Robinson, d | .000 | .038 | .075 | .112 | .014 | .014 | .014 | .014 |
| RRA, H | .578 | .605 | .631 | .658 | .012 | .012 | .013 | .013 |
| FA(1), H | .480 | .519 | .559 | .599 | .037 | .039 | .040 | .042 |
| FA(2), H | .491 | .530 | .569 | .609 | .036 | .037 | .038 | .040 |
| FA(3), H | .488 | .527 | .566 | .605 | .037 | .038 | .039 | .041 |

Notes

Each result is based on 5,000 replications of a simulated NIID or self-affine returns series. The RRA is described by equations [5] to [7]. The FA is described by [8] to [11]. The GPH and Robinson estimators are described by [12] and [13], respectively.

Table 7 Mean and standard deviation of various estimators of the Hurst exponent, L-stable process

| Estimator | Mean | | | | Standard deviation | | | |
|---|---|---|---|---|---|---|---|---|
| Parameter values | $\alpha$=2 H=0.50 | $\alpha$=1.85 H=0.54 | $\alpha$=1.72 H=0.58 | $\alpha$=1.61 H=0.62 | $\alpha$=2 H=0.50 | $\alpha$=1.85 H=0.54 | $\alpha$=1.72 H=0.58 | $\alpha$=1.61 H=0.62 |
| Sample, T=2000 | | | | | | | | |
| Pickands, H | -.279 | -.157 | -.029 | .109 | .176 | .178 | .184 | .187 |
| Hill, H | .212 | .318 | .412 | .498 | .018 | .036 | .046 | .055 |
| HR, H | .199 | .455 | .556 | .634 | .021 | .147 | .158 | .172 |
| FA(1), H | .477 | .514 | .550 | .586 | .050 | .058 | .066 | .075 |
| Parameter values | $\alpha$=2 H=0.50 | $\alpha$=1.85 H=0.54 | $\alpha$=1.72 H=0.58 | $\alpha$=1.61 H=0.62 | $\alpha$=2 H=0.50 | $\alpha$=1.85 H=0.54 | $\alpha$=1.72 H=0.58 | $\alpha$=1.61 H=0.62 |
| Sample, T=5000 | | | | | | | | |
| Pickands, H | -.275 | -.162 | -.033 | .106 | .110 | .113 | .112 | .116 |
| Hill, H | .212 | .318 | .413 | .498 | .011 | .023 | .030 | .034 |
| HR, H | .145 | .467 | .561 | .630 | .015 | .123 | .132 | .144 |
| FA(1), H | .479 | .516 | .554 | .591 | .037 | .046 | .055 | .061 |

Notes

Each result is based on 5,000 replications of a simulated NIID or self-affine returns series. The FA is described by equations [8] to [11]. The Pickands, Hill and HR estimators are described by [14].

Table 8 Summary descriptive statistics: daily logarithmic returns for 11 stock market indices

| | Mean | Standard deviation | Skewness | Kurtosis |
|---|---|---|---|---|
| Nikkei | -.00014 | .0149 | -0.28 | 11.42 |
| FTSE | .00015 | .0114 | -0.52 | 13.56 |
| SP500 | .00024 | .0119 | -1.37 | 33.23 |
| OMX Helsinki | .00027 | .0169 | -0.39 | 11.88 |
| CAC | .00016 | .0139 | -0.14 | 8.92 |
| DAX | .00015 | .0125 | -0.31 | 9.62 |
| ISEQ | .00011 | .0126 | -0.83 | 14.66 |
| MIBTel | .00008 | .0121 | -0.48 | 7.47 |
| AEX | .00016 | .0140 | -0.26 | 11.85 |
| Madrid SE | .00022 | .0128 | -0.11 | 11.10 |
| OMX Stockholm | .00033 | .0134 | -0.00 | 8.87 |

.

Table 9 Estimation results: 11 stock market indices

| Method | RRA | FA(1) | FA(2) | FA(3) | Robinson | Hill |
|---|---|---|---|---|---|---|
| Parameter | H | H | H | H | d | H |
| *Unfiltered* | | | | | | |
| Nikkei | .569* | .543* | .506 | .460 | -.036 | .327 |
| FTSE | .560 | .532 | .473 | .396 | .003 | .345 |
| SP500 | .543 | .564*** | .505 | .429 | -.042 | .383 |
| OMX Helsinki | .619*** | .608*** | .580*** | .547* | .029 | .345 |
| CAC | .561 | .547* | .518 | .476 | -.015 | .346 |
| DAX | .596** | .588*** | .537 | .484 | .016 | .347 |
| ISEQ | .614** | .594*** | .575** | .542 | .066 | .390 |
| MIBTel | .617** | .557* | .542 | .516 | .082 | .301 |
| AEX | .584 | .584** | .537 | .482 | .006 | .404 |
| Madrid SE | .603** | .558* | .524 | .474 | .019 | .343 |
| OMX Stockholm | .597*** | .615*** | .559** | .497 | .029 | .372 |
| *Filtered* | | | | | | |
| Nikkei | .582 | .546 | .506 | .456 | -.008 | .330 |
| FTSE | .575 | .506 | .472 | .404 | .013 | .342 |
| SP500 | .568 | .522 | .505 | .448 | .013 | .364 |
| OMX Helsinki | .615*** | .592*** | .573** | .538 | .014 | .351 |
| CAC | .579 | .545 | .525 | .490 | -.009 | .342 |
| DAX | .592** | .576** | .533 | .486 | .008 | .336 |
| ISEQ | .595*** | .574** | .563** | .534 | .020 | .402 |
| MIBTel | .590** | .547 | .530 | .505 | .004 | .311 |
| AEX | .577 | .553* | .518 | .473 | -.003 | .400 |
| Madrid SE | .588** | .541 | .509 | .461 | -.013 | .348 |
| OMX Stockholm | .601*** | .595*** | .554* | .500 | .017 | .370 |

Notes to Table 9

For the tests of $H_0$:H=0.5 (or d=0) against $H_1$:H>0.5 (or d>0) based on the RRA, FA and Robinson estimators, *** denotes rejection of $H_0$ in favour of $H_1$ at the 0.01 significance level. ** and * denote rejection at the 0.05 and 0.1 levels, respectively. For the test based on the Hill estimator, $H_0$:H=0.5 would be rejected at the 0.01 significance level in every case.